\documentclass[acus]{JAC2003}

\usepackage{graphicx}
\usepackage{booktabs}

\begin{document}

\title{Transmission Efficiency Measurement at the FNAL 4-rod 
RFQ\thanks{Work supported by Fermilab Research Alliance, LLC under 
Contract No. DE-AC02-07CH11359 with the United States Department of 
Energy.}}

\author{J.-P. Carneiro\thanks{carneiro@fnal.gov}, F. G. Garcia, J.-F. Ostiguy, A. Saini, R. Zwaska, Fermilab, Batavia, IL 60510\\
        B. Mustapha, P. Ostroumov, ANL, Argonne, IL 60439}

\maketitle

\begin{abstract}
This paper presents measurements of the beam transmission performed on the 4-rod RFQ currently under operation at Fermilab. 
The beam current has been measured at the RFQ exit as a function of the magnetic field strength in the two LEBT solenoids. 
This measurement is compared with scans performed on the FermiGrid with the beam dynamics code \texttt{TRACK}. 
A particular attention is given to the impact, on the RFQ beam transmission, of the space-charge neutralization 
in the LEBT.
\end{abstract}

\section{INTRODUCTION}
A new injector has been in operation since 2012 on the
FNAL 400 MeV Linac as a part of the Proton Improvement Plan 
whose primary goal is to increase the proton flux in the booster to ultimately 2.25$\times$10$^{17}$ protons per hour. 
This new injector, composed of an ion source, a Low Energy Beam Transport line (LEBT), a 4-rod Radio Frequency Quadrupole (RFQ) and a Medium Energy Beam Transport (MEBT) is presented in details in Ref.\cite{cytan}. The beam transmission in the injector, from the ion source to the MEBT exit, has routinely been measured since the start of its operation ranging from 
40\% to 50\%. This measured transmission is significantly lower than the expected one, which according to computer simulations should be close to 100\%. 
After a brief description of the injector, this paper presents a measurement of a beam transmission at the MEBT exit as a function of the LEBT solenoid fields. This measurement is compared to numerical simulations from the code \texttt{TRACK}~\cite{track}. The simulations reveal that the space charge neutralization pattern, which is unlikely to be homogeneous along the LEBT, plays a crucial role in the injector transmission.

\section{THE FNAL LINAC INJECTOR}

\begin{figure}[t]
\centering
\includegraphics*[width=1.0\columnwidth]{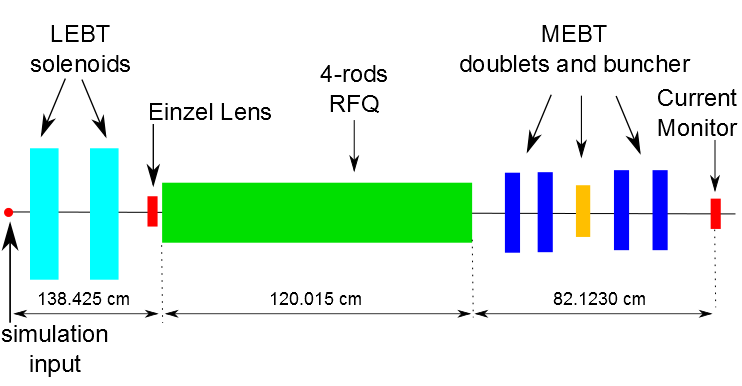}
\caption{Layout of the FNAL Linac Injector.}
\label{fig01}
\end{figure}

 A layout of the injector is depicted in Fig.~\ref{fig01}. A magnetron source produces 35~keV H$^{-}$ bunches of typically  100~$\mu$s long, at a repetition rate of 15~Hz with an average current ranging from 50 to 70~mA. The LEBT comprises two solenoids that match the H$^{-}$ beam produced by the source into the RFQ entrance. The beam is further accelerated to $750$~keV by a 4-rod RFQ operating at 201.25~MHz. At the exit of the RFQ, the MEBT matches the beam into the first Drift-Tube Linac Tank (DTL\#01) using two doublets for transverse matching and an RF buncher operating at 3~MV/m for longitudinal matching. An Einzel lens operating at -37.7~kV is installed near the entrance of the RFQ to chop the first 20~$\mu$s of the pulse. A current monitor which is used in the measurement described in the next session is located~8.25~cm from the downstream face of the last MEBT quadrupole, i.e at the DTL\#01 entrance. The total length of the injector line is in the order of 4 meters. 

\section{TRANSMISSION MEASUREMENT}
\begin{figure}[h]
\centering
\includegraphics*[width=1.0\columnwidth]{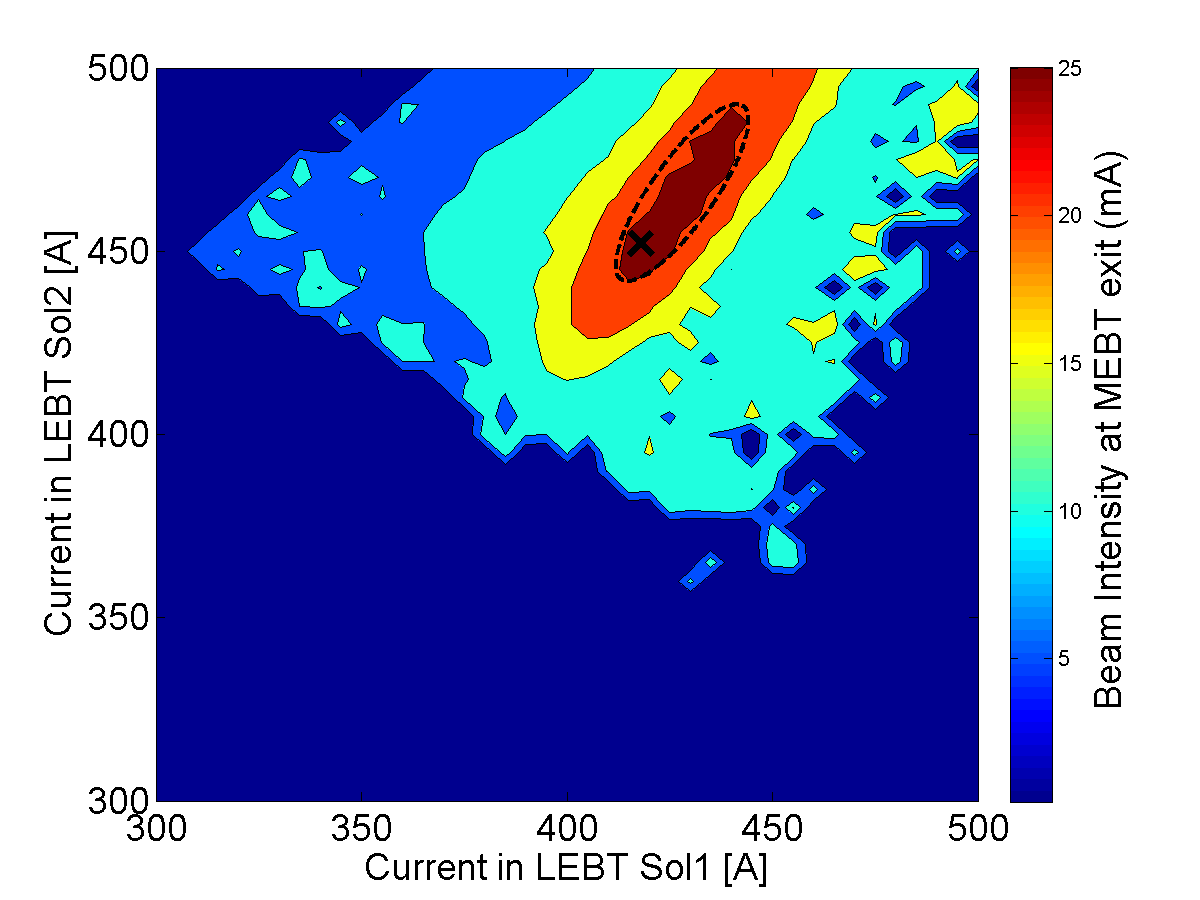}
\caption{Measurement of the beam intensity at the current monitor located at the exit of the MEBT as a function of the LEBT solenoid currents. The cross represents the actual point of operation of the LEBT and the ellipse mimics the area of favorable transmission.}
\label{fig02}
\end{figure}

An experiment has been performed on the linac injector which consisted in varying, in a systematic way, the current in the two LEBT solenoids and measuring the beam intensity on the current monitor located at the MEBT exit. This measurement is represented in Fig.~\ref{fig02}.

For this measurement, the LEBT solenoids have been scanned from 300~A to 500~A with a step of 5~A, all other parameters in the injector being kept at their optimal values. During the measurement, the vacuum on the LEBT was measured to be 4$\times$10$^{-6}$~Torr using a cold cathode gauge located in the middle of the LEBT. The maximum beam intensity measured at the MEBT exit and reported in Fig.~\ref{fig02} is 27.3~mA for a measured current at the ion source of 47.5~mA, which represents a transmission of 57.5\%.

\section{BEAM DYNAMICS SIMULATION}
An attempt to numerically reproduce, with the beam dynamics code \texttt{TRACK}, the experimental scan presented in Fig.~\ref{fig02}
has been undertaken at Fermilab. The code \texttt{TRACK}, developed at ANL, has been selected for this work because of its ability to use an external 3D field map to simulate an RFQ. The EM fields were extracted from a MicroWave Studio model of the 4-rod RFQ built by S. Kurennoy at LANL. The advantage of using 3D fields to simulate the RFQ lies on the fact that, as discussed in \cite{kurennoy}, the 4-rod RFQ presents a field asymmetry due to the stems which can lead to emittance increase and beam losses along the RFQ. 
A script launches 10201 \texttt{TRACK} runs on the FermiGrid, 
representing a scan of the two LEBT solenoids from 0 to 500~A, with 5~A steps. Another script analyses for each run the predicted beam transmission at the position of the current monitor in the MEBT exit.  
A 4D Waterbag has been used as input distribution with 5$\times$10$^{4}$ macro-particles, 3D fields maps have been generated for the LEBT solenoids and the MEBT buncher has been modeled using a simulated axial electric field. The radius of aperture has been kept at 5~cm all along the LEBT but in the last 8~cm where it was reduced to 2.2~cm to take into account the Einzel lens aperture. In the \texttt{TRACK} model, a 1~cm aperture hole at the RFQ entrance was assumed. The vane aperture has been implemented in the code along the RFQ and in the MEBT the radius of aperture has been kept at 2~cm.

\begin{figure}
\centering
\includegraphics[width=1.0\columnwidth]{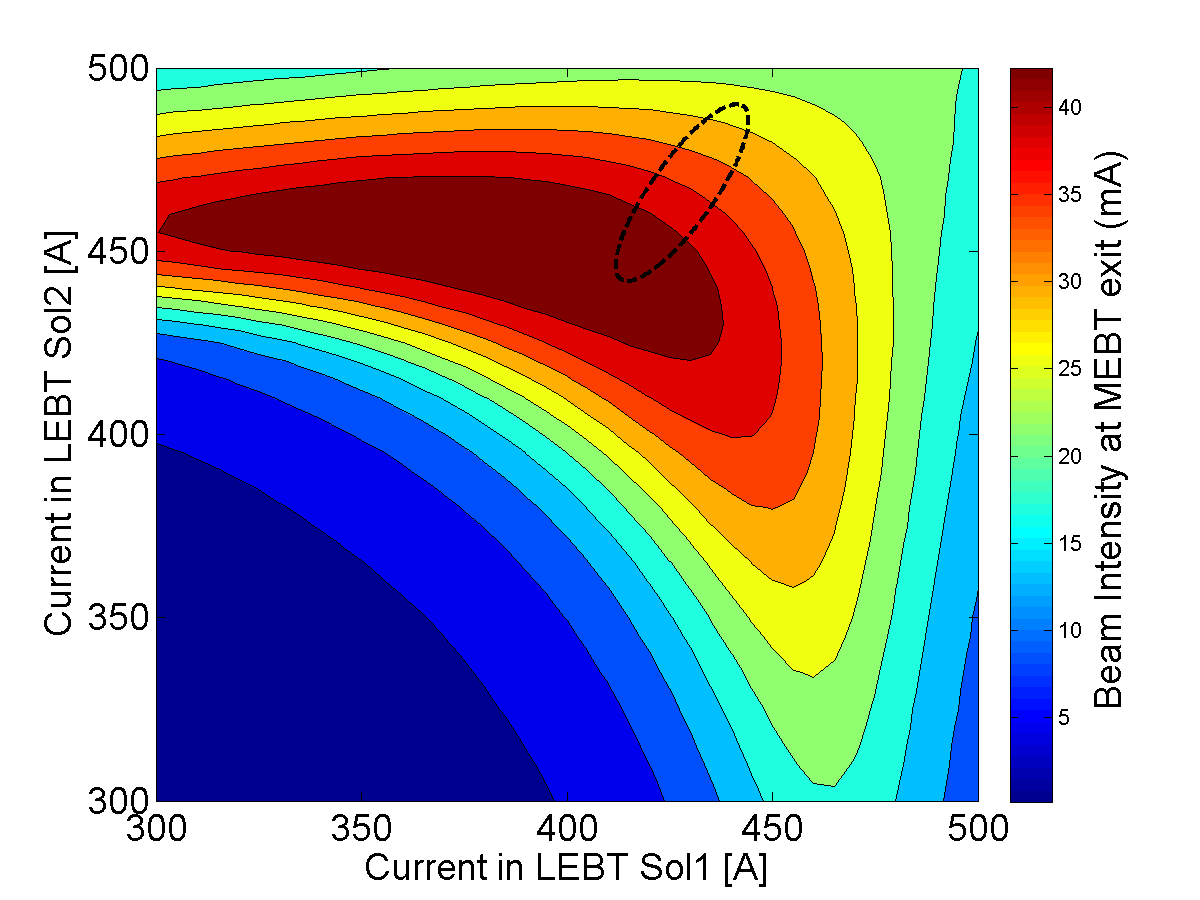} 
\put(-230,170){$\textbf{(a)}$} \quad
\qquad
\includegraphics[width=1.0\columnwidth]{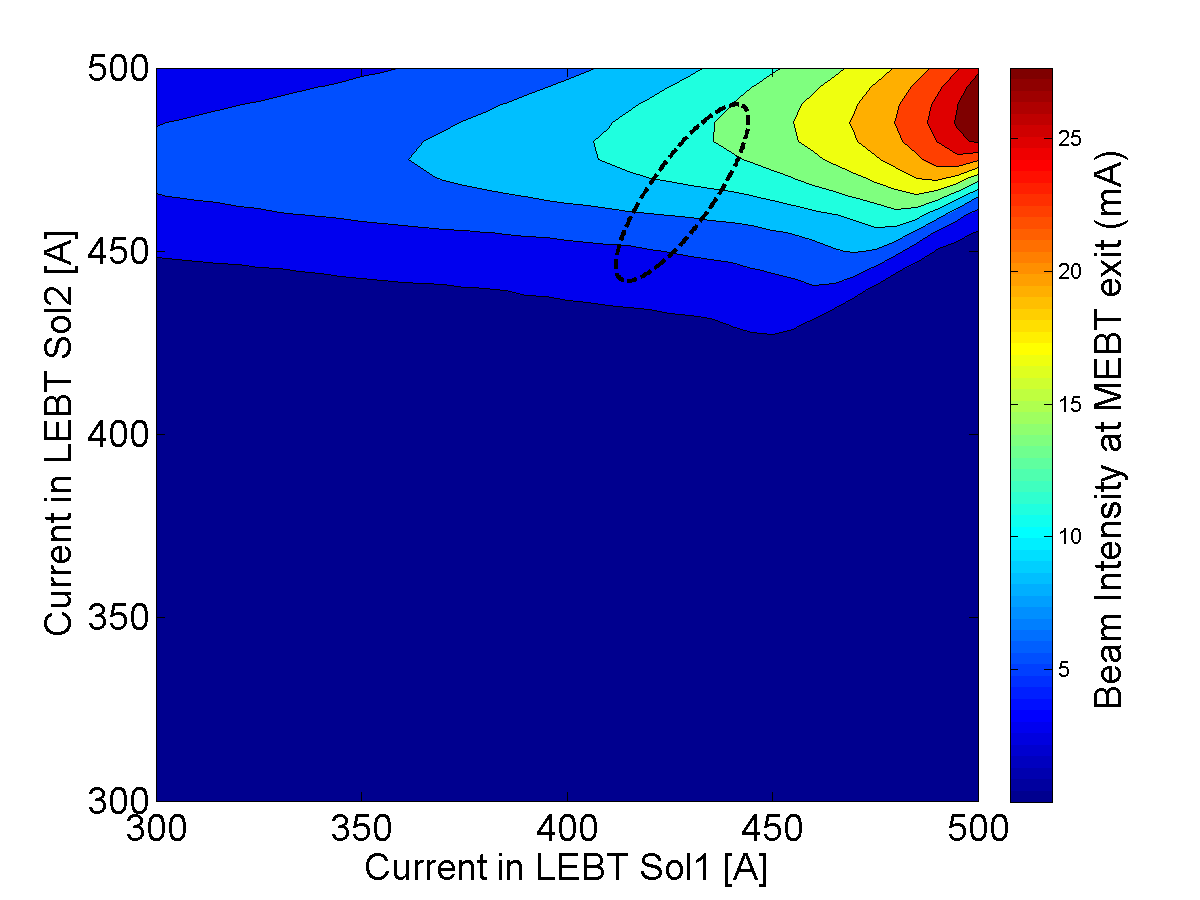}
\put(-230,170){$\textbf{(b)}$} 
\caption{Simulated beam intensity at the MEBT exit as a function of the LEBT solenoid currents for a uniform neutralization factor along the LEBT of (a) 100\% and (b) 60\%. From \texttt{TRACK}. The ellipse mimics the measured area of favorable transmission reported in Fig.~\ref{fig02}.}
\label{fig03}
\end{figure} 

\subsection*{LEBT neutralization}
An important parameter to take into account in the simulations is the space charge neutralization factor in the LEBT. In fact, upon exciting the ion source, the H$^{-}$ beam interacts with the 
H$_{2}$ molecules present in the residual gas creating H$_{2}^{+}$ ions and electrons. The H$_{2}^{+}$ ions are then trapped in the beam potential well and counteract the beam space charge field while the electrons are ejected to the beam pipe wall. In \texttt{TRACK} the neutralization factor is modeled by a simple reduction of the beam intensity. We considered in our simulation three scenarios: a uniform space charge neutralization factor in the LEBT, a region at the entrance of the RFQ which is un-neutralized and an hypothetical space charge neutralization pattern along the LEBT.  

\subsection*{Uniform LEBT neutralization factor}
We performed with \texttt{TRACK} a scan of the LEBT solenoid currents for a uniform neutralization pattern along the LEBT (from the simulation starting point to the RFQ entrance hole) ranging from 0\% (full space charge) to 100\% (no space charge), with steps of 10\% in the neutralization factor. Figure \ref{fig03}(a) and \ref{fig03}(b) 
show respectively the cases 100\% and 60\%, where the simulated beam current is reported at the location of the current monitor at the MEBT exit. In these figures is also reported the ellipse which depicts the region of favorable transmission measured in Fig.~\ref{fig02}. Clearly, the agreement between the simulated and measured area of favorable transmission is poor for these two cases as for all others not reported in this document. Our conclusion from these studies is that a model based on uniform neutralization is too simplistic. The neutralization is, in 
fact, unlikely to be uniform.

\begin{figure}[t]
\centering
\includegraphics*[width=0.9\columnwidth]{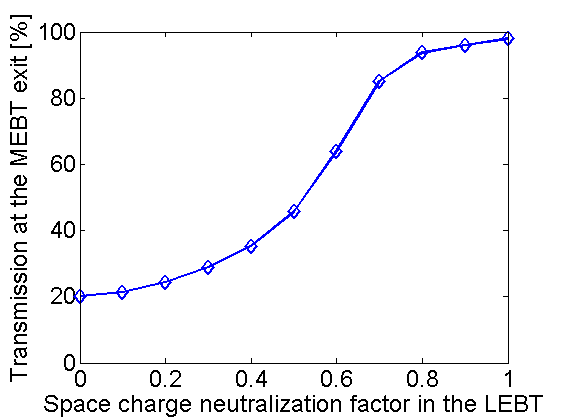}
\caption{Maximum simulated transmission at the MEBT exit as a function of a uniform neutralization factor along the LEBT. From \texttt{TRACK}.}
\label{fig04}
\end{figure}

Figure~\ref{fig04} reports for each neutralization factor scan above-mentioned the maximum transmission predicted by \texttt{TRACK}.  
We believe that, as the neutralization factor decreases, the beam develops along the LEBT non linear space charge effects which inhibit a proper matching into the RFQ resulting in beam losses. For a beam fully neutralized, \texttt{TRACK} predicts a beam transmission of 98\%  which drops to 60\% (close to the measured value) for a beam neutralized 
at 60\%.

\vspace{-16mm}
\subsection*{Un-neutralized region at the LEBT end}
\begin{figure}[h]
\centering
\includegraphics*[width=0.9\columnwidth]{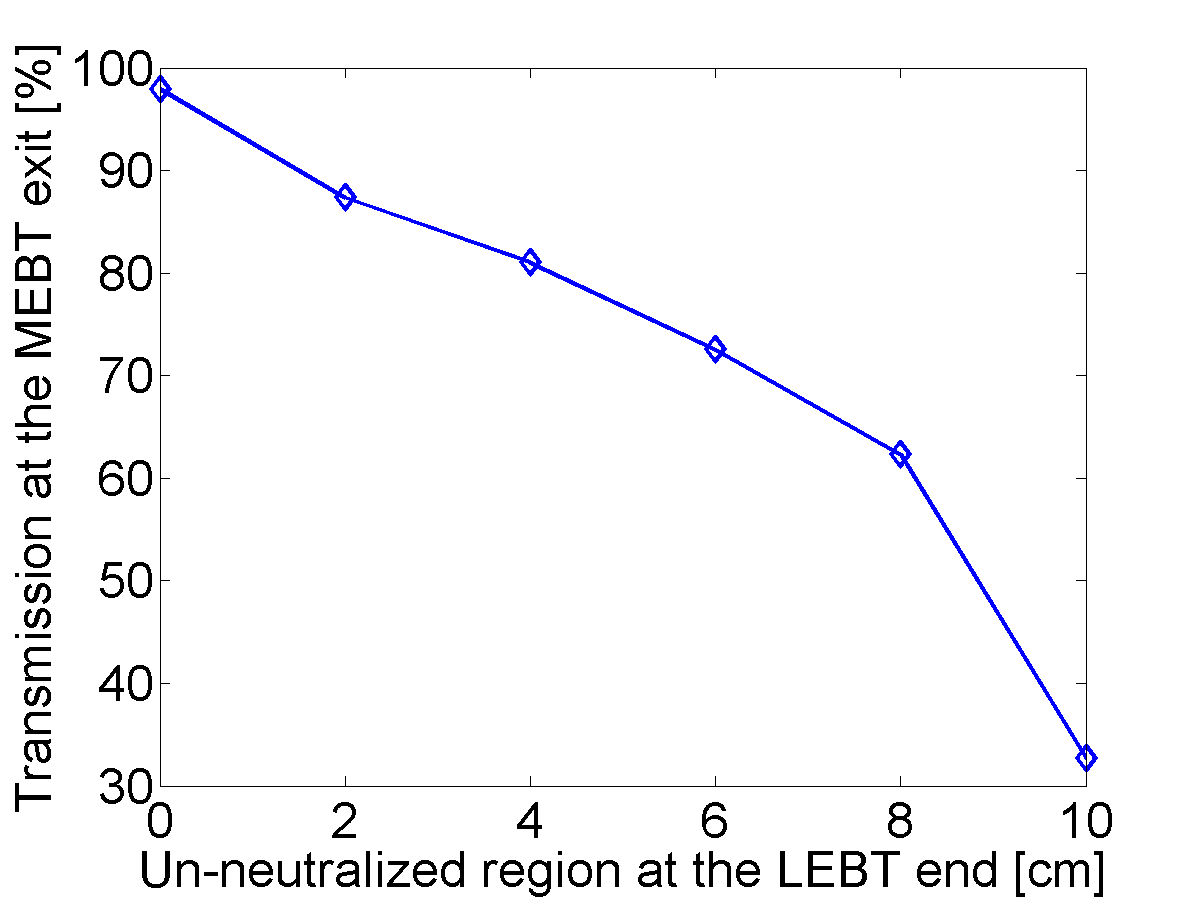}  
\caption{Maximum simulated transmission at the MEBT exit as a function of an un-neutralized region at the LEBT end, the remaining part neutralized at 100\%. From \texttt{TRACK}.}
\label{fig05}
\end{figure}

We performed another set of LEBT solenoid scans with \texttt{TRACK} taking a 100\% fully neutralized beam all along the LEBT but in the last few centimeters at the LEBT end. Fig.~\ref{fig05} shows that few centimeters of un-neutralized region at the LEBT end has a significant impact in the beam transmission. We could speculate that the level of neutralization is inversely correlated with the beam size and that at this region the neutralization could be less effective. For instance, if we consider the LEBT 100\% neutralized but the last 8~cm, the beam transmission at the MEBT exit drops to 63\%. 
\subsection*{Non-uniform LEBT neutralization factor}
\begin{figure}[h]
\centering
\includegraphics[width=1.0\columnwidth]{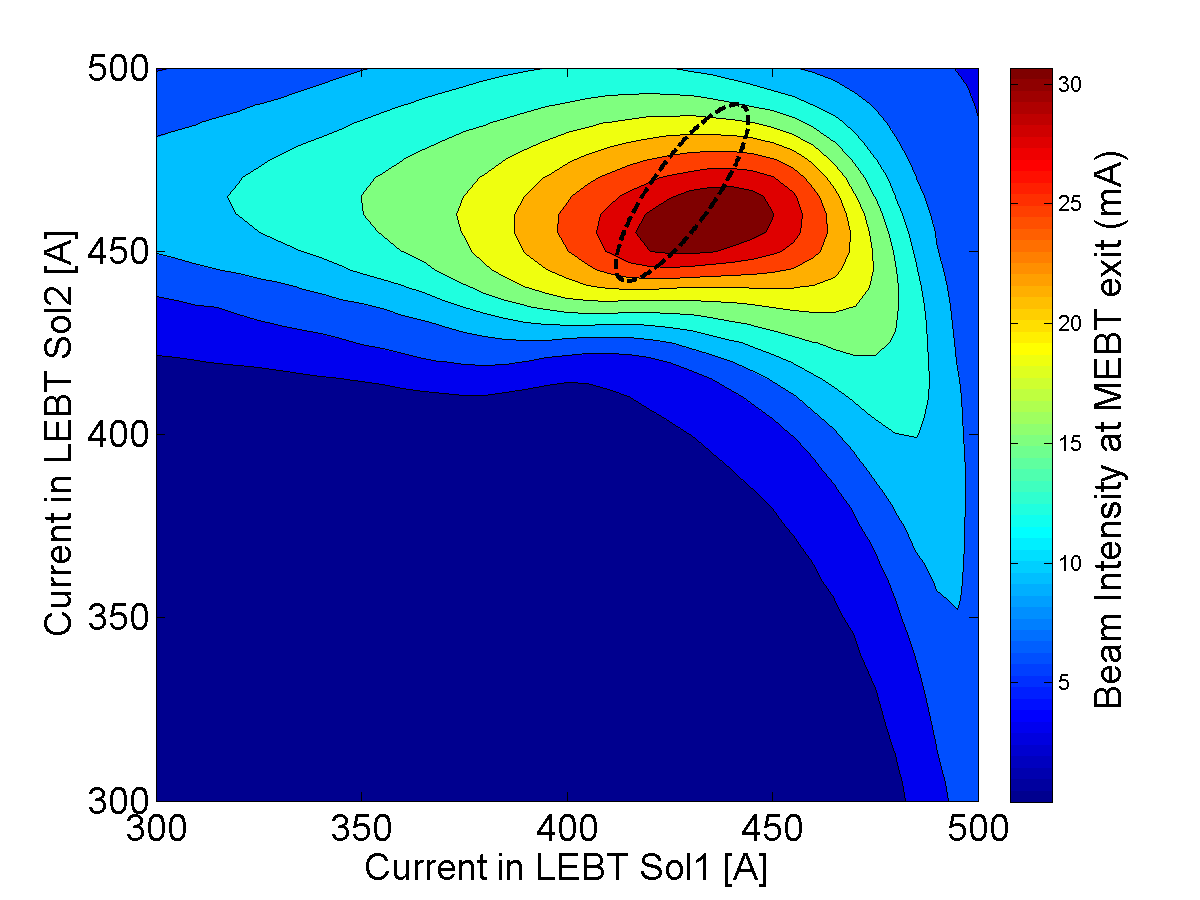} 
\put(-230,170){$\textbf{(a)}$}  \quad
\qquad
\includegraphics[width=1.0\columnwidth]{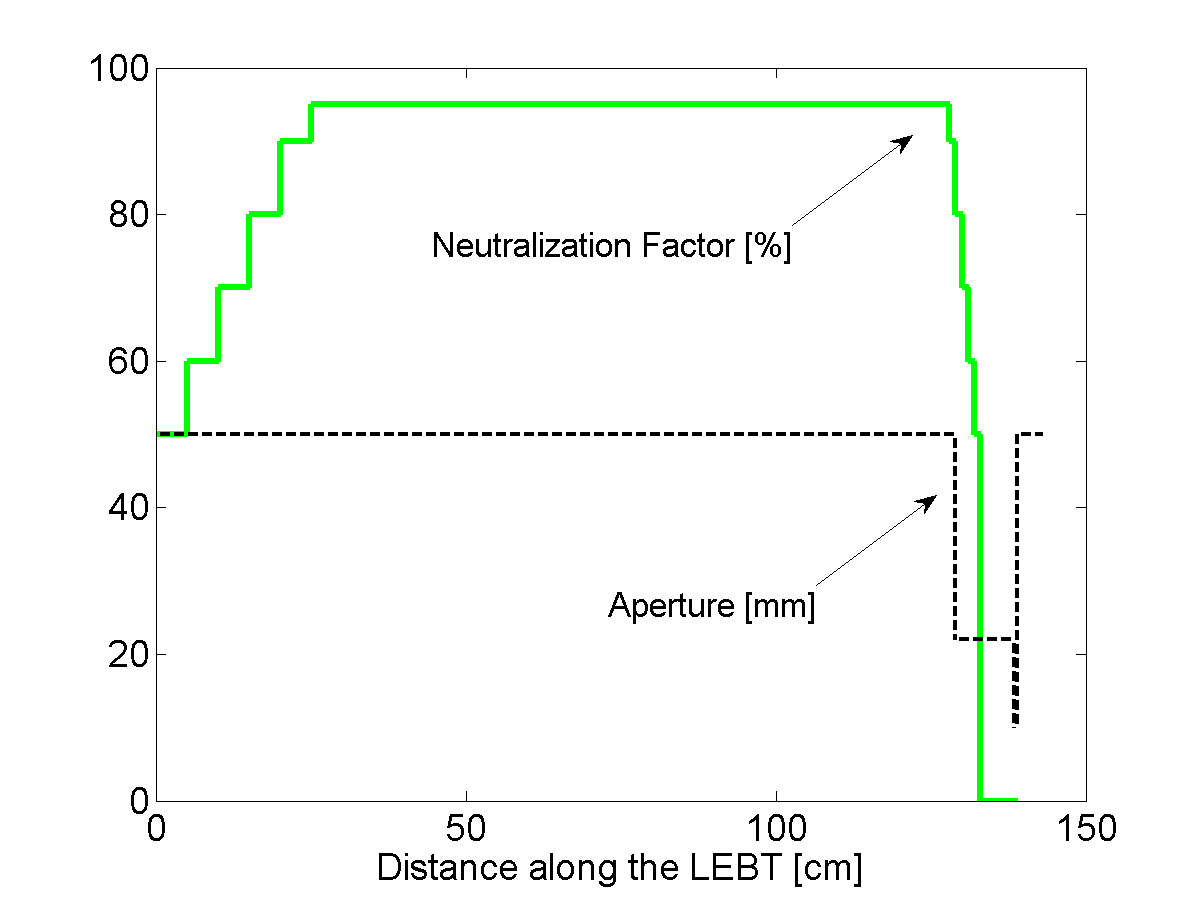} 
\put(-230,170){$\textbf{(b)}$}  
\caption{(a) Simulated beam intensity at the MEBT exit as a function of the LEBT solenoid currents for (b) a non-uniform neutralization along the LEBT. From \texttt{TRACK}. The ellipse mimics the measured area of favorable transmission reported in Fig.~\ref{fig02}.}
\label{fig06}
\end{figure}

Figure~\ref{fig06}(a) shows a scan performed with \texttt{TRACK} 
taking a non-uniform neutralization factor along the LEBT, as depicted in Fig.~\ref{fig06}(b). We considered a neutralization profile with gradual linear increase and decrease respectively at the upstream and downstream extremities of the LEBT. The last 6~cm of the LEBT were considered un-neutralized. With this pattern, the simulations presented in Fig.~\ref{fig06}(a) show better agreement with the measurement presented in Fig.~\ref{fig02}, particularly concerning the size and location of the areas of favorable beam transmission. Yet, the predicted transmission from \texttt{TRACK} is 20\% higher than the measured one.
\vspace{-4mm}
\vspace{-8mm}
\section{CONCLUSION}
The simulations results indicate that it is likely that the neutralization profile in the LEBT can account for the measured transmission. More work will be needed in order to understand and better characterize the neutralization pattern.
\vspace{-4mm}
\section{ACKNOWLEDGEMENT}
The author would like to thank C.Y Tan for sharing information about the injector, B. Pellico and V. Shiltsev for continuing support.   

\end{document}